\documentstyle[12pt,epsf]{article}
\newcommand{\bfig}{\begin{figure}}
\newcommand{\efig}{\end{figure}}

\newcommand{\pdot}{{\displaystyle{\raisebox{-1.5ex}[0.25ex]{$\cdot$}
     \atop\raisebox{0.6ex}[0.25ex]{$\scriptstyle (p)$}}}}
\newcommand{\prdot}{{\displaystyle{\raisebox{-1.5ex}[0.25ex]{$\cdot$}
     \atop\raisebox{0.6ex}[0.25ex]{$\scriptstyle (p^{\prime})$}}}}
\newcommand{\ppdot}{{\displaystyle{\raisebox{-1.5ex}[0.25ex]{$\cdot$}
     \atop\raisebox{0.6ex}[0.25ex]{$\scriptstyle (p,p^{\prime})$}}}}

\newcommand{\kuroten}{{\displaystyle{\raisebox{-1.5ex}{$\odot$}
     \atop\raisebox{0.3ex}[1.25ex]{$\scriptstyle(p,p^{\prime})$}}}}
\newcommand{\peke}{{\displaystyle{\raisebox{-1.5ex}{$\times$}
      \atop\raisebox{0.3ex}[1.25ex]{$\scriptstyle (p,p^{\prime})$}}}}

\def\ap{\alpha^{\prime}}

\def\12{\frac{1}{2}}
\def\bea{\begin{eqnarray}}
\def\eea{\end{eqnarray}}
\def\ba{\begin{array}}
\def\ea{\end{array}}

\def\one-loop{\mbox{\scriptsize one-loop}}

\def\G{\Gamma}

\catcode`\@=11

\@addtoreset{equation}{section}
\def\theequation{\arabic{section}.\arabic{equation}}

\def\@normalsize{\@setsize\normalsize{15pt}\xiipt\@xiipt
\abovedisplayskip 14pt plus3pt minus3pt%
\belowdisplayskip \abovedisplayskip
\abovedisplayshortskip  \z@ plus3pt%
\belowdisplayshortskip  7pt plus3.5pt minus0pt}

\def\small{\@setsize\small{13.6pt}\xipt\@xipt
\abovedisplayskip 13pt plus3pt minus3pt%
\belowdisplayskip \abovedisplayskip
\abovedisplayshortskip  \z@ plus3pt%
\belowdisplayshortskip  7pt plus3.5pt minus0pt
\def\@listi{\parsep 4.5pt plus 2pt minus 1pt
            \itemsep \parsep
            \topsep 9pt plus 3pt minus 3pt}}

\def\underline#1{\relax\ifmmode\@@underline#1\else
        $\@@underline{\hbox{#1}}$\relax\fi}
\@twosidetrue

\relax

\catcode`@=12

\catcode`\@=11

\def\section{\@startsection{section}{1}{\z@}{3.5ex plus 1ex minus
   .2ex}{2.3ex plus .2ex}{\large\bf}}

\def\thesection{\Roman{section}.}

\def\appendix{\setcounter{section}{0}
        \def\thesection{Appendix }
        \def\theequation{\Alph{section}.\arabic{equation}}}

\def\ps@headings{\def\@oddfoot{}\def\@evenfoot{}
\def\@oddhead{\hbox{}\hfill
        \makebox[.5\textwidth]{\raggedright\ignorespaces --\thepage{}--
        \hfill {}}}
\def\@oddhead{\hbox{}\hfill --\thepage{}-- \hfill
        {}}
\def\@evenhead{\@oddhead}
\def\subsectionmark##1{\markboth{##1}{}}
}

\ps@headings

\catcode`\@=12

\relax

\def\figcap{\section*{Figure Captions\markboth
        {FIGURECAPTIONS}{FIGURECAPTIONS}}\list
        {Fig. \arabic{enumi}:\hfill}{\settowidth\labelwidth{Fig. 999:}
        \leftmargin\labelwidth
        \advance\leftmargin\labelsep\usecounter{enumi}}}
 \relax
\def\tablecap{\section*{Table Captions\markboth
        {TABLECAPTIONS}{TABLECAPTIONS}}\list
        {Table \arabic{enumi}:\hfill}{\settowidth\labelwidth{Table 999:}
        \leftmargin\labelwidth
        \advance\leftmargin\labelsep\usecounter{enumi}}}
 \relax
\def\reflist{\section*{References\markboth
        {REFLIST}{REFLIST}}\list
        {[\arabic{enumi}]\hfill}{\settowidth\labelwidth{[999]}
        \leftmargin\labelwidth
        \advance\leftmargin\labelsep\usecounter{enumi}}}
 \relax

\catcode`\@=11

\def\ps@headings{\def\@oddfoot{}\def\@evenfoot{}
\def\@oddhead{\hbox{}\hfill
        \makebox[.5\textwidth]{\raggedright\ignorespaces --\thepage{}--
        \hfill {}}}
\def\@evenhead{\@oddhead}
\def\subsectionmark##1{\markboth{##1}{}}
}

\ps@headings

\relax

\newskip\humongous \humongous=0pt plus 1000pt minus 1000pt

\newif\ifdtup

\def\beq{\begin{equation}}
\def\eeq{\end{equation}}

\def\beqn{\begin{eqnarray}}
\def\eeqn{\end{eqnarray}}
\relax

\def\G2{{\; \rm GeV/}c^2}
\def\G{\; \rm GeV}

\def\dotx{\dotx{\dot\overline{x}}}

\relax

\begin{document}

%
%
\begin{titlepage}

\renewcommand{\thefootnote}{\fnsymbol{footnote}}

\begin{flushright}
      \normalsize     
    OU-HET 361 \\  October, 2000 \\ 
         hep-th/0010066  \\
\end{flushright}

%
\begin{center}
  {\large\bf Correspondence between Noncommutative Soliton \\
 and Open String/D-brane System via Gaussian Damping Factor}%
\footnote{This work is supported in part
 by the Grant-in-Aid  for Scientific Research
(12640272, 12014210) from the Ministry of Education,
Science and Culture, Japan,
and in part by
the Japan Society for the Promotion of Science for Young
Scientists.}
\end{center}

\vfill

\begin{center}
    { {  B. Chen}\footnote{e-mail address:
                                 chenb@funpth.phys.sci.osaka-u.ac.jp},
      { H. Itoyama}\footnote{e-mail address:
                               itoyama@funpth.phys.sci.osaka-u.ac.jp},
      { T. Matsuo}\footnote{e-mail address:
                                matsuo@funpth.phys.sci.osaka-u.ac.jp}
      and { K. Murakami}\footnote{e-mail address:
                              murakami@funpth.phys.sci.osaka-u.ac.jp}}\\
\end{center}

\vfill

\begin{center}
      \it  Department of Physics,
        Graduate School of Science, Osaka University,\\
        Toyonaka, Osaka 560-0043, Japan
\end{center}

\vfill


\begin{abstract}
The gaussian damping factor (g.d.f.) and the new interaction vertex with
  the symplectic tensor are the characteristic properties 
 of the $N$-point scalar-vector scattering amplitudes of the $p-p^\prime
  (p < p^\prime)$ open string system which realizes noncommutative geometry. 
The g.d.f. is here interpreted as a form factor of the $Dp$-brane 
by noncommutative $U(1)$ current. Observing that the g.d.f. is in fact
equal to the Fourier transform
of the noncommutative projector soliton introduced by Gopakumar, 
Minwalla and Strominger, we further identify the $Dp$-brane in
 the zero slope limit with
the noncommutative soliton state.
It is shown that the g.d.f. depends only on the total momentum of
$N-2$ incoming/outgoing photons in the zero slope limit. In the description of
the low-energy effective action (LEEA) proposed before,
 this is shown to 
follow  from the delta function propagator and the form of the initial/final
wave functions in the soliton sector which resides in 
$x^{m} \; (m= p+1 \cdots p^{\prime})$
 dependent part of the scalar field $\Phi(x^\mu, x^m)$.  
 The three and four point amplitudes computed from LEEA
 agree with string calculation.   We discuss related issues
  which are resummation/lifting of infinite degeneracy and
 conservation of momentum transverse to the $Dp$-brane.

\end{abstract}

\vfill

\setcounter{footnote}{0}
\renewcommand{\thefootnote}{\arabic{footnote}}

\end{titlepage}


\section{Introduction}

  Notion of $D$-brane has led people to think what string theory
  ought to be beyond perturbation theory. Investigations
  of its spacetime properties are, however, so far limited to
 classical solutions to supergravity theory and its connection to
  configurations of branes. One reason to prevent a more direct study
 at string amplitudes is that, in the zero-slope limit,
  this object becomes too singular to study and gets simply removed
  from the low energy physics except that the momenta carried by the modes
 of a string are constrained.  In an interesting setup of
  string theory with constant $B_{MN}$ background \cite{Callan1} 
realizing noncommutative
  geometry,  it has been found \cite{SW} that distances at all scales
 can be kept finite.
 This offers a possibility to establish direct correspondence between
  string theory in the zero slope limit and the attendant local field
  theory:
   this time  $D$-brane is present in both sides as physical degrees of
 freedom. We will accomplish
  this correspondence in an open string connecting a $Dp$-brane and
    a $Dp^{\prime}$-brane with the $Dp$-brane inside, 
following the series of work
   \cite{CIMM, CIMM2} which has uncovered  a number of properties: these
  include $1)$ spectrum which contains a large number of light states and
  $2)$ the appearance in string amplitudes of a symplectic tensor $J$
  and a multiplicative factor decaying exponentially with momenta 
 (a gaussian damping factor). These are derived from
  several nontrivial worldsheet properties of system.
 
  In quite different vein, it has been argued  \cite{GMS} that
 classical soliton solution is
  possible to construct in scalar noncommutative field theory, 
avoiding the no-go theorem of Derrick.
 We will find  that the soliton solution of this type, in particular,
 the simplest projector soliton is just the right representation of the
 $Dp$-brane
 in field theory side  in order to establish the correspondence.

 For definiteness, let us first specify the process studied in this paper.
At an initial state, we prepare a $Dp$-brane which is at rest and which lies
in the worldvolume of a $Dp^{\prime}$-brane. The $Dp^{\prime}$-brane
is regarded as entire space in this paper.
  We place the tachyon (the lowest mode) of a $p-p^{\prime}$ open
 string which carries a
 momentum $k_{1\mu},$ $\mu = 0\cdots p$
along the $Dp$-brane worldvolume. In addition, $N-2$ noncommutative 
$U(1)$ photons carrying momenta $k_{a M}$ 
$a=3\cdots N,\; M= 0 \cdots p^{\prime}$ in $p^{\prime} +1$ dimensions are 
present.  They get absorbed into the $Dp$-brane.
 At a final state, the $Dp$-brane
is found to be present and the momentum of the tachyon is measured to be
 $-k_{2\mu}$ along the $Dp$-brane worldvolume.  We will examine the tree
 scattering amplitude of this process both from string perturbation theory of
 the D-brane/open string system in the zero slope limit
 and from perturbation theory of the field theory action proposed
 in \cite{CIMM2}.  We will find that computations from both sides in fact
 agree by identifying the $Dp$-brane with an initial/final configuration
  representing a noncommutative soliton.   

 The computation of the scattering amplitude of this process from string
 perturbation theory has been already carried out in \cite{CIMM2}.
  In the next section, we will begin with recapitulating its properties,
 focusing upon the gaussian damping factor (g.d.f.) which is originally
 associated with each external vector leg. In the zero slope limit, the
 desirable cross terms develop
 and the g.d.f. is shown to be an overall multiplicative factor for any $N$
  which depends only upon the total momentum.
 An approximate resummation of infinitely many light states \cite{CIMM}
 propagating in the $t$-channel \cite{CIMM2}
 is responsible for this phenomenon, which we will refer to as lifting
 of the infinite degeneracy.
  Finally in this section, we observe \cite{OBS} that the g.d.f. is in
 fact equal to
 the Fourier transform of the noncommutative projector soliton solution
 introduced by Gopakumar, Minwalla and Strominger in  \cite{GMS}.
The g.d.f. is naturally interpreted as a form factor of the $Dp$-brane
by noncommutative $U(1)$ current. The $Dp$-brane in the zero slope limit is
identifed with the noncommutative soliton state.

In section three, we consolidate this identification and interpretation
in the light of the low energy effective action (LEEA) which is proposed
in \cite{CIMM2}. The adequate description of the process above  is given
 by perturbation theory
  of this LEEA which at the same time permits us to define a soliton
 sector residing
 in the  $x^{m}$ $(m= p+1 \cdots p^{\prime})$ dependent part of the scalar
 field  $\Phi(x^\mu, x^m)$.  
  That the g.d.f. depends on the total photon momentum alone is
 found to be a simple consequence  from the delta function propagator
 in perturbation theory and the
  form of the initial/final wave function given by Fourier transform of
 the projector
  soliton solution.  The three and four point tree amplitudes agree 
 with string calculation. It is satisfying to see that string theory realizing
 noncommutative geometry and the attendant local field theory in fact share
 the several interesting properties  which are derived from two
 completely different lines of reasoning.
 Section four is devoted to outlook and a few comments which are 
 more speculative. 
  We basically follow the notation of \cite{CIMM2}. With regard to
 the spacetime index,
 $M,N \cdots$ run from $0$ to $p^{\prime}$, $\mu, \nu \cdots$ from
 $0$ to $p$ and   $m,n \cdots$  from $p+1$ to $p^{\prime}$.

\section{Gaussian damping factor of the scattering amplitude from
the $p-p^\prime$ open string with constant $B_{ij}$ field}

Let us recall how the gaussian damping factor has been found in 
\cite{CIMM2} from the scattering amplitude which involves two scalars and 
$(N-2)$ massless vectors which are noncommutative $U(1)$ photons. We begin
 with the integral representation of this amplitude. 
(See \cite{CIMM2} for its derivation and more complete explanation
of  our notation.) The $SL(2, \bf{R})$ invariant integral
(Koba-Nielsen) representation of this amplitude is
\begin{eqnarray}
\lefteqn{ {\cal A}_N=c (2\pi)^{p+1}\prod^p_{\mu=0}
\delta\left(\sum^N_{e=1}k_{e\mu}\right) \int \prod^N_{a=4}dx_a
\prod^N_{a^\prime=3}d\theta_{a^\prime}d\eta_{a^\prime}
\exp{\cal C}_{a^\prime}( \left\{  \nu_I \right\} )} \nonumber\\
& &\times \prod^N_{c=4}\left[
          x_c^{-\alpha^\prime s_c+\alpha^{\prime} m^2_T}
          (1-x_c)^{2\alpha^{\prime} k_3\pdot k_c}\right]
    \prod_{4\leq c<c^\prime\leq N}
       (x_c-x_{c^\prime})^{2\alpha^{\prime} k_{c^\prime}\pdot k_c}
\nonumber \\
& &\times \prod_{3\leq c<c^\prime\leq N}
   \exp\left[-2\alpha^{\prime}\sum_{I,\overline{J}}G^{I\overline{J}}
     \left\{ \kappa_{cI}\overline{\kappa}_{c^\prime\overline{J}}
            {\cal H}\left(\nu_I; \frac{x_c}{x_{c^\prime}}\right)
           +\overline{\kappa}_{c\overline{J}} \kappa_{c^{\prime}I}
          {\cal H}\left(\nu_{I};\frac{x_{c^{\prime}}}{x_{c}}\right)
     \right\} \right] 
\nonumber\\
& &\times \exp\left({\rm NC}\right)
 \exp\Big([0,2]+[2,0]+[1,1]+[2,2]\Big)
\Bigg|_{x_{1}=0,x_{2}=\infty,x_3=1}~.\label{Koba}
\end{eqnarray}
Here we have used the $SL(2, \bf{R})$ invariance to fix the locations of 
two tachyon vertex operators and  that of a massless vector vertex operator
to be respectively at $x_1=0, x_2=\infty,$  and $x_3=1$
($x_a\equiv -\xi_a=e^{\tau_a}$).  From now on, we set $c= 2i$ and
 multiply the expression by
   $  \left( - \sqrt{2 \alpha^{\prime}} \right)^{N-4}$.
 We explain eq. (\ref{Koba})
further:
\begin{enumerate}
\item We have employed the worldsheet superfield formalism. 
Eq. (\ref{Koba}) involves integrations over fermionic
 variables $\theta_a$ and $\eta_a$.
The terms containing $\theta_a$ and $\eta_a$ are classified  by
 the number of $\eta_a$ and by the number of $\theta_a$, which we
 designate respectively
 by the first and by the second entry inside the bracket. These are given as
 $[0,2], [2,0], [1,1]$, and $[2,2]$.
The explicit forms of these terms  can be found in \cite{CIMM2}. ( See also
 \cite{IM} for a comparison to the more familiar $Dp-Dp$ case with
 vanishing $B$.)

\item The term denoted by $\exp({\rm NC})$ originates from the
 noncommutativity of the worldvolume. It extends into all directions of
 $Dp^{\prime}$-brane worldvolume:
\begin{eqnarray}
({\rm NC})
 =\sum_{1\leq a<a^{\prime}\leq N}\frac{i}{2}
      \epsilon (x_{a}-x_{a^{\prime}})\sum_{M ,N=0}^{p^{\prime}}
      \theta^{MN}k_{aM}k_{a^{\prime}N}~,
\label{eq:realnc}
\end{eqnarray}
where $k_{1m}=k_{2m}=0$ for $m=p+1,\ldots,p^{\prime}$, and
 $\theta^{2A-1,2A} = \frac{2 \pi \alpha^{\prime} b_{A}}
{ \varepsilon(1+b^2_A)}$
  is the noncommutativity parameter.

\item The momentum dependent multiplicative factor 
$\exp {\cal C}_{a}(\left\{  \nu_I \right\}  )$  with
\begin{equation}
{\cal C}_{a}( \left\{  \nu_I \right\} )=\alpha^{\prime}\sum_{I,\overline{J}}
    2 \kappa_{aI}\overline{\kappa}_{a\overline{J}} G^{I\overline{J}}   
   \left\{ \gamma + \frac{1}{2}\Big(\mbox{\boldmath${\psi}$}(\nu_{I})
            +\mbox{\boldmath${\psi}$}(1-\nu_{I})\Big)\right\}~
\label{eq:C} 
\end{equation}
comes from the subtracted Green function at a coincident point. 
 This Green function has been introduced in \cite{CIMM2} and is defined
 by the difference of the two distinct Green functions, namely, the one
 defined with respect to the $SL(2, \bf{R})$ invariant vacuum and
 the other with respect to
  the oscillator vacuum.
  In eq. (\ref{eq:C}), $\gamma, \psi(\nu_{I})$
are the Euler constant and the digamma function respectively,
 and $G^{I\overline{J}}$ is the inverse of the open string metric for
 the $x^{p+1}, \cdots,
 x^{p^\prime}$ directions
\begin{equation}
G^{I\overline{J}}=G^{\overline{J}I}=\frac{2}{\varepsilon(1+b^2_I)}
\delta^{I\overline{J}}~.
\end{equation}
We denote by $\kappa_{I}, \overline{\kappa}_{\overline{J}}$ a set of
 momenta in complex
 notation defined as 
\begin{equation}
\label{eq:kappaandk}
\kappa_{I} = \frac{1}{2}\left(k_{2I-1}-ik_{2I}\right)~,
\hspace{4ex}
\overline{\kappa}_{\overline{I}}
 =\frac{1}{2}\left(k_{2I-1}+ik_{2I}\right)~.
\end{equation}

\item There are a few different ways in which an inner product of two
      vectors $A_i$  and  $B_j$ is taken with respect to the open string
 metric.  These are denoted 
by $\pdot$, $\prdot$ and $\ppdot$, depending on the directions concerned with. 
  For example,
\beqn
A\pdot B  &=& \sum^p_{\sigma,\rho=0}G^{\sigma\rho}A_{\sigma}
 B_{\rho} \;\;,  \;\;\;{\rm and}   \nonumber \\
k \ppdot \zeta   
  &=&  \sum_{I,\overline{J}=\frac{p+2}{2}}^{\frac{p^{\prime}}{2}}
  \left(G^{I\overline{J}}\kappa_{I}\overline{e}_{\overline{J}}
        +G^{\overline{J}I}\overline{\kappa}_{\overline{J}}e_{I}\right)
=\sum_{I,\overline{J}=\frac{p+2}{2}}^{\frac{p^{\prime}}{2}}
   G^{I\overline{J}}\left(\kappa_{I}\overline{e}_{\overline{J}}
                           +\overline{\kappa}_{\overline{J}}e_{I}
     \right) \;\;.
\eeqn
  We also introduce
$\kuroten$ and $\peke$  to denote the "incomplete inner products"
\begin{equation}
 \left(k \kuroten \zeta\right)_{I}=\sum_{\overline{J}}
G^{I\overline{J}}(\kappa_{I}\overline{e}_{\overline{J}}
                  +\overline{\kappa}_{\overline{J}}e_{I})~,
\quad
\left(k \peke \zeta \right)_{I}=\sum_{\overline{J}}
  \frac{2\delta^{I\overline{J}}(\kappa_{I}\overline{e}_{\overline{J}}
        {}-\overline{\kappa}_{\overline{J}}e_{I})}
       {\varepsilon (1+b_{I}^{2})}~,
\end{equation}
  so that
\begin{equation}
 \sum_{I}\left(k\kuroten\zeta\right)_{I}
  =k\ppdot\zeta~,
\quad
 \sum_{I}\left(k\peke \zeta\right)_{I}
 =ik\ppdot J\zeta~.
\end{equation}
  Here the matrix $J$ is a $(p^{\prime}+1)\times (p^{\prime}+1)$
antisymmetric matrix defined as
\begin{equation}
  J =\left( {J_{\mu}}^{\rho}\right)\equiv 
   \begin{array}{r@{}l}
    \left(\begin{array}{ccc|ccccc}
	 0&      & &  & &      & & \\
          &\ddots& &  & &      & & \\
          &      &0&  & &      & & \\ \hline
          &      & &0 &1&      & & \\
          &      & &-1&0&      & & \\
          &      & &  & &\ddots& & \\
          &      & &  & &      &0&1\\
          &      & &  & &      &-1&0
        \end{array}\right)
             &   \begin{array}{l}
                 \mbox{\scriptsize $0$} \\  \vdots\\ 
                 \mbox{\scriptsize $p$}\\ \mbox{\scriptsize $p+1$}\\
                 \mbox{\scriptsize $p+2$}\\  \vdots\\
                 \mbox{\scriptsize $p^{\prime}-1$}\\
                 \mbox{\scriptsize $p^{\prime}$}
                 \end{array}        
   \end{array}~.
\end{equation}

\item The functions ${\cal H}(\nu_I;\frac{x_c}{x_{c^\prime}})$ or  
${\cal H}(\nu_I;\frac{x_{c^\prime}}{x_c})$ is defined by the 
hypergeometric series as 
\begin{equation}
{\cal H}(\nu; z) =\left\{\begin{array}{lll}
   \displaystyle{\cal F} \left(1-\nu ;\frac{1}{z}\right)
     {}-\frac{\pi}{2} b
    &=  \displaystyle{ \sum_{n=0}^{\infty}\frac{z^{-n-1+\nu }}{n+1-\nu} }
     {}-\frac{\pi}{2} b  &
     \mbox{ for $|z| > 1$}\\[3ex]
  \displaystyle{\cal F} \left( \nu ; z \right)  \;\;\;\;+\frac{\pi}{2} b
  &=   \displaystyle{ \sum_{n=0}^{\infty}\frac{z^{n+ \nu }}{n + \nu} } \;\;
   +\frac{\pi}{2}b &
     \mbox{ for $|z| <1$} \;\;.
\end{array}
\right.  \label{eq:h-prop}
\end{equation}
The two infinite series in the above defining relation
should be analytically continued to each other.

\end{enumerate}

 {} From now on, we will focus on the nontrivial zero slope limit of the
 amplitude.
The zero slope limit is defined as
\begin{equation}
\begin{array}{rcl}
\alpha' &\sim& \varepsilon^{1/2} \to 0~,  \\
g &\sim& \varepsilon \to 0~, \\
|b_{I}| &\sim& \varepsilon^{-1/2} \to \infty~.
\end{array} \end{equation} 
 This limit keeps $\alpha' b_{I}$ finite:
\beq
\alpha' b_{I} \to \beta_{I}\;\;\;.
\label{eq:psib}
\eeq
  In terms of the open string metric and the noncommutativity parameter,  
this implies
\beq
      \frac{1}{2\pi} \left( J G \theta \right)_{2I-1}^{\;\;2I-1} =
      \frac{1}{2\pi} \left( J G \theta \right)_{2I}^{\;\;2I}  = \beta_{I} \;\;.
\eeq
 In addition, the following limit is taken without loss of generality:
\beq
\label{eq:signb}
\nu\equiv\nu_{\frac{p+2}{2}}  \rightarrow  1  \;, \;\;\;\;
\nu_{\widetilde{I}}  \rightarrow 0  \;, \;\;    {\rm for} \;\;
 \widetilde{I}\neq \frac{p+2}{2}\;\;,
\eeq
 so that
\beq
\label{eq:signb2}
 b_{\frac{p+2}{2}} \to  + \infty \;\;, \;\;\;
 b_{\widetilde{I}} \to  - \infty \;\;.
\eeq

It is simple to obtain the three-point amplitude:
\begin{equation}
  {\cal A}_{3}= -i (2\pi)^{p+1}\prod_{\mu=0}^{p}
  \delta\left(\sum^3_{a=1}k_{a\mu}\right)
\left\{(k_2-k_1)\pdot \zeta_3-ik_3\ppdot J \zeta_3
\right\}  e^{{\cal C}_{3}(  \left\{  \nu_I \right\} )}
e^{\frac{i}{2}\theta^{ij}k_{1i}k_{2j}}~.
\label{eq:A3}
\end{equation}
The first term is the derivative coupling of a charged scalar with a
 massless vector and vanishes when there is no momentum transfer of
 the tachyon in the directions of $Dp$-brane worldvolume.
The second term is the term found in \cite{CIMM2}. Both
are multiplied by the term representing noncommutativity 
$e^{\frac{i}{2}\theta^{ij}k_{1i}k_{2j}}$ as well as by
the factor $\exp {\cal C}(\left\{  \nu_I \right\})$.

This exponential multiplicative factor
 $\exp {\cal C}(\left\{  \nu_I \right\})$
defined in eq.~(\ref{eq:C}) becomes in the zero slope limit
\begin{equation}
  \exp{\cal C}( \left\{  \nu_I \right\} )\to  \exp \left(
-\pi\sum_{I,\bar{J}}\left|\beta_{I}\right|
\kappa_{I}\overline{\kappa}_{\overline{J}}G^{I\overline{J}}  \right)
=  \exp   \left( -\frac{\pi}{2}\sum_{I}\left|\beta_{I}\right|
  \left(k\kuroten k\right)_{I}  \right)
  \equiv D \left(k_{m} \right)
\;\;, \label{eq:damping2}
\end{equation}
 where we have used
$\mbox{\boldmath$\psi$}(1)= -\gamma$ and
 eqs.~(\ref{eq:kappaandk}), (\ref{eq:psib}). 
This factor is associated with each vector propagating into the 
$x^{p+1} \sim x^{p^{\prime}}$ directions. We will refer to this as
 gaussian damping factor (g.d.f.) in the rest of this paper.
 Inserting
 the initial and final wave functions
 of the tachyon and  that of the noncommutative $U(1)$ photon,
(which we should have inserted in the first place together with
 the vertex operators,)
 we obtain
\beqn
  \lim  {\cal A}_{3}= -i (2\pi)^{p+1}\prod_{\mu=0}^{p}
  \delta\left(\sum^3_{a=1}k_{a\mu}\right) 
 \frac{1}{(2\pi)^{p/2}}\frac{1}{\sqrt{2\omega_{{\vec k}_{2}}}}
\frac{1}{(2\pi)^{p/2}}\frac{1}{\sqrt{2\omega_{{\vec k}_{1}}}}
\frac{1}{(2\pi)^{p^\prime/2}}
\frac{1}{\sqrt{2|{\vec k}_{3}|}}   \;\; \nonumber \\
  \times \left\{(k_2-k_1)\pdot \zeta_3-ik_3\ppdot J \zeta_3
\right\}  D(k_{3m})
  e^{\frac{i}{2}\theta^{ij}k_{1i}k_{2j}} \;\;.
\label{eq:lima3}
\eeqn

 Let us come back to eqs. (\ref{eq:signb}), (\ref{eq:signb2}).
It has been observed that, due to this fine tuning of the sign of
 $b_I$,  a large number of light states appear in the limit. To be
 more  precise, these light states are obtained by acting
 the several low-lying fermionic modes on  the oscillator
 vacuum  and  multiplying by an
 arbitrary polynomial consisting of the lowest bosonic mode.
   This latter bosonic mode is the one which has failed
to become a momentum due to the boundary condition of the $p-p^{\prime}$
open string and is responsible for an infinite number of nearly degenerate
 low-lying states. We will see that the string amplitude in fact has resummed
 and lifted this approximate infinite degeneracy by evaluating
  its effect as an exponential factor
 and that this lifting renders
the net g.d.f. of the amplitude to depend only upon the total momentum of the
incoming photons.

   In order to prove this last statement, let us recall that the
 contributions to
 the $N$-point amplitude surviving the zero
 slope limit come from the endpoints of the $N-3$ integrations: 
at the endpoints
  $x_{c}$  coalesces to either $x_{c-1}$ or $x_{c+1}$ and eventually
 gets close to either $0$ or $1$. We need only to analyse the behavior
 of  the function ${\cal H}(\nu;x)$ or 
${\cal H}(\nu; \frac{1}{x})$ on the exponent near the $x=0$ and
 the $x=1$, paying attention
to the order of the integrations and the limit.
\noindent
 In the region  $\frac{x_{c^{\prime}}}{x_{c}} \approx 0$ for $c^{\prime}>c,$ 
we find that the factor
\beqn
\label{eq:factor}
  P_{c^{\prime} c} \equiv -2\alpha^{\prime}
\sum_{I,\overline{J}}G^{I\overline{J}}
     \left( \kappa_{cI}\overline{\kappa}_{c^\prime\overline{J}}
         {\cal H}\left(\nu_I; \frac{x_c}{x_{c^\prime}}\right)
        +\overline{\kappa}_{c\overline{J}} \kappa_{c^{\prime}I}
     {\cal H}\left(\nu_{I};\frac{x_{c^{\prime}}}{x_{c}}\right) \right) \:\:
\eeqn
can be approximated by
\beqn
\label{eq:appx1}
  P_{c^\prime c} & \approx & -\pi 
\sum_I|\beta_I|(k_c\kuroten k_{c^\prime})_I \nonumber \\
& &-\alpha^\prime\sum_I\Big((k_c\kuroten k_{c^\prime})
+(k_c\peke k_{c^\prime})\Big)_I
\lim \frac{(\frac{x_{c^\prime}}{x_c})^{1-\nu_I}-1}{1-\nu_I} \nonumber \\
& &-\alpha^\prime\sum_I\Big((k_c\kuroten k_{c^\prime})
-(k_c\peke k_{c^\prime})\Big)_I
\lim \frac{(\frac{x_{c^\prime}}{x_c})^{\nu_I}-1}{\nu_I} \;\;. 
\eeqn
 Here we have exploited that the contribution to 
${\cal H}(\nu; \frac{1}{x})$ from
 the modes other than the lowest bosonic mode is ignorable in the limit
  and that
the constant piece is safe to evaluate in the limit. The limit in the
 last two terms
 of eq. (\ref{eq:appx1})
should be taken after the integration and will be discussed in the next
 paragraph.
\noindent
On the other hand, the factor $P_{c^{\prime} c}$  in the region
 $\frac{x_{c^{\prime}}}{x_{c}} \approx 1$
can be evaluated, using the expansion of ${\cal H}(\nu; \frac{1}{x})$
 near $x=1$.
 We find that $P_{c^{\prime} c}$ can be approximated by
\beqn
\label{eq:appx2}
  P_{c^\prime c}  \approx  -\pi \sum_I
| \beta_I|(k_c\kuroten k_{c^\prime})_I 
+2\ap \sum_I (k_c\kuroten k_{c^\prime})_I 
\log(1-\frac{x_{c^\prime}}{x_c}) \;\;.
\eeqn

 We see that in either region the factor $P_{c^{\prime} c}$ contains
 the identical  constant piece 
$\displaystyle{-\pi \sum_I|\beta_I|(k_c\kuroten k_{c^\prime})_I }$ in the
 limit.   Multiplying   
$\displaystyle{ \prod_{3\leq c<c^\prime\leq N} \exp \left[
 -\pi \sum_I \mid  \beta_I \mid (k_c\kuroten k_{c^\prime})_I \right] }$ by
$\displaystyle{\prod^N_{a=3} D(k_{am})}$,
 (  see eq. (\ref{eq:damping2})), we find that the amplitude $A_{N}$ contains
 an overall multiplicative factor
\begin{equation}
 D  \left( \sum_{a=3}^{N} k_{am} \right) = 
 \exp\left[-\frac{\pi}{2}\sum_{I}\left|\beta_{I}\right|
  \left((  \sum_{a=3}^{N} k_{a})\kuroten 
(\sum_{b=3}^{N} k_{b})\right)_{I}\right]\;\;,
\label{eq:damping}
\end{equation} 
 which depends upon the total photon momentum alone.
 This is what we wanted to show.

 The singular behavior near   $\frac{x_{c^{\prime}}}{x_{c}} \approx 1$
 of the second term of eq. (\ref{eq:appx2}) on the exponent of 
 $\exp P_{c^{\prime} c}$
 is responsible for the massless pole in the $s$-channel.
It is instructive therefore to estimate near 
$\frac{x_{c^{\prime}}}{x_{c}} \approx 0$
 the second and the third terms of eq. (\ref{eq:appx1})  on the exponent of
 $\exp P_{c^{\prime} c}$, which we discuss qualitatively here.  Expanding 
 $\exp P_{c^{\prime} c}$ in Taylor series,
combining with the other factors in the integrand and integrating over
 $x_{c^{\prime}}$ near the origin, we find propagators of an infinite number
 of states corresponding to the light spectrum of
 a $p-p^{\prime}$ open string in
  the $t$-channel.
 We see that, in the treatment of eq. (\ref{eq:appx1}), we have taken care of
this approximate infinite degeneracy on the exponent and
 its principal contribution
is the cross term of the net g.d.f.  It is this resummation of spectrum
 of states coming from the lowest bosonic  mode that has provided
 this cross term.
 The infinite degeneracy has been lifted. A closer look at eq. (5.8) of
 \cite{CIMM2} 
 shows that ignoring the second and the third terms in eq. (\ref{eq:appx1})
  corresponds to setting to zero the mass differences among
 the infinitely many light states
  due to the lowest bosonic mode.

  Let us turn to the four point amplitude in the zero slope limit. 
After lifting the infinite degeneracy due to the lowest bosonic mode, 
we still  have the contributions from several nearly degenerate states due
 to the lowest fermionic modes. 
 In the subsequent section, we will discuss 
 only those parts of the amplitude in which
the state with the lowest mass (tachyon) participates.  
See eq. (5.8) of \cite{CIMM2} 
 for the complete formula in the zero slope limit  which contains the above 
 contributions as well.
Taking into account what we have established above, we find
\begin{eqnarray}
\lim   {\cal A}_{4}&=& -2i (2\pi)^{p+1}
\prod_{\mu=0}^{p}\delta \left(\sum_{a=1}^{4}k_{a\mu}\right)
D \left( k_{3m} + k_{4m} \right) \;
   \exp\left(\frac{i}{2}\theta^{\mu\nu}k_{1\mu}k_{2\nu}
              +\frac{i}{2}\theta^{MN}k_{3M}k_{4N}\right) 
\nonumber\\
 &&  \frac{1}{(2\pi)^{p/2}}\frac{1}{\sqrt{2\omega_{{\vec k}_{2}}}}
\frac{1}{(2\pi)^{p/2}}\frac{1}{\sqrt{2\omega_{{\vec k}_{1}}}}
\frac{1}{(2\pi)^{p^\prime/2}}\frac{1}{\sqrt{2|{\vec k}_{3}|}}
\frac{1}{(2\pi)^{p^\prime/2}}\frac{1}{\sqrt{2|{\vec k}_{4}|}}
    \nonumber\\
 && \left[ \frac{1}{t-m^{2}}
\frac{1}{2}\left\{\left(k_{2}-(k_{1}+k_{4})\right) \pdot \zeta_{3}
  {}-ik_{3}\ppdot J\zeta_{3} \right\} \right.\nonumber\\
&& \hspace{6em} \left\{\left((k_{2}+k_{3})-k_{1}\right) \pdot
     \zeta_{4} -ik_{4}\ppdot J\zeta_{4} \right\} \nonumber\\
&&\hspace{1em}  +\frac{1}{s}\left\{
   \left(
   (k_{2}-k_{1})\pdot\zeta_{3}-i(k_{3}+k_{4})\ppdot J\zeta_{3}
  \right) k_{3}\prdot\zeta_{4}\right.\nonumber\\
&&\hspace{4em}-\left(
       (k_{2}-k_{1})\pdot\zeta_{4}
     {}-i(k_{3}+k_{4})\ppdot J\zeta_{4} \right)
     k_{4}\prdot\zeta_{3}  \nonumber\\
&&\hspace{4em}
   + \left( \frac{1}{2}(k_{3}-k_{4})\pdot (k_{1}-k_{2})
    {}-ik_{3}\ppdot J k_{4}
    \right)    \zeta_{3}\prdot\zeta_{4}\bigg\} \bigg]
   \nonumber\\
&&  +\left(k_{3}\leftrightarrow k_{4};
   \zeta_{3}\leftrightarrow\zeta_{4}\right)
   \;\;.
\label{eq:st4ampstring}
\end{eqnarray}

 We end this section by giving an explicit connection  and  related
 remarks between the g.d.f. and
 the noncommutative projector soliton, which is a key observation 
we make to the remainder of
 this paper.
Let us first rewrite the g.d.f. as 
\begin{eqnarray}
 D(k_{m})    &=&\exp  \left( -\frac{1}{4}
\sum^{\frac{p^\prime}{2}}_{I=\frac{p+2}{2}}
 \mid \theta^{2I-1,2I}  \mid (k_{2I-1}k_{2I-1}+k_{2I}k_{2I}) 
 \right) \nonumber\\
&=&\prod^{\frac{p^\prime}{2}}_{I=\frac{p+2}{2}}
\tilde{\phi_0}(k_{2I-1},k_{2I} ;\theta^{2I-1,2I}) \;\;.
\end{eqnarray}
Observe that
\begin{eqnarray}
\label{eq:observe}
  2 \pi \mid \theta \mid
\tilde{\phi_0}&=&\int d^2x e^{ik_1x^1+ik_2x^2}
\phi_0(x^1,x^2 ;\theta) \;\;,  \nonumber \\
\phi_0(x^1,x^2 ;\theta)&=&
2e^{-\frac{1}{ \mid \theta \mid}((x^1)^2+(x^2)^2)} \;\;.
\end{eqnarray} 
Function $\phi_0(x^1,x^2 ;\theta)$ is the projector soliton solution of the
 noncommutative scalar field theory discussed in \cite{GMS}. 
It satisfies $\phi_0\ast \phi_0=\phi_0$ and  is
represented as a ground state projector $|0\rangle\langle0|$ in the 
Fock space representation of noncommutative algebra $[x^1, x^2]=i\theta$.
In \cite{GMS}, $\phi_0$ is discussed as a soliton solution of noncommutative 
scalar field theory in the large $\theta$ limit.  In our discussion,
 Fourier transform of $\phi_0$  is seen to appear for all values of $\theta$.

  There are a few points associated with this observation.
 From our discussion, it is natural to interpret that some of
 the physical degrees
 of freedom of the $Dp$-brane are participating in the process although
 the $D$-brane
 is introduced in the first quantized string through boundary conditions.
  Eq. (\ref{eq:lima3}) tells us rather obviously that 
the g.d.f.  $D(k_{3m})$ is a form factor of the $Dp$-brane
 by noncommutative $U(1)$ current, which can be written as
\beq
  \left(  \Phi^\dagger   \stackrel{\leftrightarrow}{\partial}^\mu \Phi ,
-i\partial_{n} \left(\Phi^\dagger J^{mn}\Phi \right)  \right)
   \;\;,
\label{eq:ncu1c}
\eeq
  using the scalar field $\Phi(x^{\mu},x^{m})$ discussed in the next section.
Putting together this fact and the obsevation  of the last paragraph, 
we identify the $Dp$-brane
 in the zero slope limit with the noncommutative soliton.
The form factor $D(k_{3m})$ can either be  Fourier transform of
 the classical profile of
 the $Dp$-brane/soliton or quantum mechanical wave function of this object
 in momentum space.
  Perturbation theory presented in the next section supports the latter
 point of view.

 Another point on the degrees of freedom of the $Dp$-brane is the issue
 regarding with
 the conservation of momenta in the directions transverse to
 the $Dp$-brane worldvolume.
 In string calculation, there is no delta function which ensures
 the conservation of
  momenta in these directions  as we place spin and twist fields on
 the worldsheet
  \cite{DFMS,GNS}. 
On the other hand, our result eq. (\ref{eq:damping}) does show
 that these momenta have been deposited in the $Dp$-brane.
 The momentum conservation holds
  for the combined system of tachyon, photons and the
 $Dp$-brane/soliton. 
The center of mass
 coordinate of  the $Dp$-brane/soliton has become activated to
 receive the photon momenta.
  One way to interpret the absence of the delta function
 is that one is measuring in string calculation
 an inclusive process  with regard to the $Dp$-brane: 
in the final state, we have
  confirmed its presence only and did not measure its momenta.

  There is an alternative way to view the scattering process not by momentum
 eigenstates but by constructing wave packets as
 initial and final configurations.
 This point of view explains the absence of the delta function 
immediately. It involves
  instead integrations over both initial and final momenta of the 
$Dp$-brane/soliton.
  We will discuss the relationship of these two points of view 
in the next section.

\section{ $Dp$-brane and the Projective Soliton of Noncommutative Scalar
 Field Theory}

We now give a field theoretic derivation of the properties of 
the string amplitude
 in the zero slope limit given by eqs. (\ref{eq:lima3}), 
(\ref{eq:damping}) and 
(\ref{eq:st4ampstring}). We will show that an adequate 
description is given in 
perturbation theory of low energy effective 
action (LEEA) proposed in \cite{CIMM2} by specifying proper
 initial and final 
states associated with the scalar field $\Phi(x^\mu, x^m)$.

In \cite{CIMM2}, the following action has been proposed:
\begin{eqnarray}
&&S = S_0 + S_1~,\nonumber\\
&&\mbox{with}\quad
S_0 = \frac{1}{g_{YM}^{\ 2}}\int d^{p'+1}x \sqrt{-G}
\left\{ -\left(D_{\mu}\Phi \right)^{\dag}\ast\left(D^{\mu}\Phi\right)
       {}-m^2 \Phi^{\dag} \ast \Phi
       {} -\frac{1}{4} F_{MN} \ast F^{MN}\right\}~,\nonumber\\
&&\hspace{3em}
S_1 = \frac{1}{2g^{\ 2}_{YM}}\int d^{p'+1}x \sqrt{-G}
\Phi^{\dag} \ast F_{mn} J^{mn} \ast \Phi~,
\label{eq:leea}
\end{eqnarray}
where 
\begin{eqnarray}
&&D_{\mu}\Phi =\partial_{\mu}\Phi-iA_{\mu} \ast \Phi~,
\qquad
\left( D_{\mu} \Phi \right)^{\dag}
  =\partial_{\mu}\Phi^{\dag}+i \Phi^{\dag}\ast A_{\mu}~,\nonumber \\
&&F_{MN}=\partial_{M}A_{N}-\partial_{N}A_{M}
    {}-i \left[A_{M},A_{N}\right]_{\ast}~,\quad
 \left[A_{M}, A_{N}\right]_{\ast}= A_{M} \ast A_{N}
    {}-A_{N} \ast A_{M} \;\;.
\end{eqnarray}
 The star product is defined by
\beq
f(x) \ast g(x)=  \left. e^{-\frac{i}{2} \theta^{MN} 
 \frac{\partial}{\partial y^{M}}
 \frac{\partial}{\partial z^{N}} }  f(y) g(z)
 \right|_{y,z \rightarrow x}   \;\;.
\eeq
  We denote by $A_M(x^\mu, x^m)$ 
 a $(p^\prime+1)$-dimensional vector
 field which corresponds to 
noncommutative $U(1)$ photon and  by $\Phi(x^\mu,x^m)$
 a  scalar field
 which corresponds
to the ground state tachyon of the $p-p^\prime$ open string with 
$\displaystyle{m^2=   - \lim_{\alpha^\prime\to 0}
   (1-  \displaystyle{ \sum_I \nu_I }) / 2\ap } $. 
Reflecting the fact that the tachyon momenta are constrained
 to lie in $p+1$ dimensions, 
the Lorentz index of the kinetic term for the scalar field
 runs from 0 to p and there
is no kinetic term for the remaining $p^\prime-p$ directions. 
 We set  $g_{YM}$  to $1$ from now on.

Perturbation theory obtained from the action (eq.(\ref{eq:leea}))
is elementary to carry out but we stop to 
explain here a few tricky points. The quantized scalar
 field $\Phi(x^\mu, x^m)$ in the
interaction picture obeys a free field equation
\beq
\label{eq:fieldeq}
(\partial_\mu\partial^\mu-m^2)\Phi(x^\mu, x^m)=0\;\;.
\eeq
Expanding $\Phi(x^\mu, x^m)$ in Fourier series, we find its mode
 expansion
\beq
\label{eq:exp1}
\Phi(x^\mu, x^m)= \frac{1}{(2\pi)^{p^\prime/2}}
\int d^pk_i\int d^{p^\prime -p}K_m 
\frac{1}{\sqrt{2\omega_{\vec k}}}
 \left( \underline{a}(k_i,K_m)e^{i\underline{k}
\prdot x}+\underline{b}^\dagger(k_i,K_m)e^{-i\underline{k}
\prdot x} \right)\;\;,
\eeq
where 
\beq
\omega_{\vec k}=\sqrt{k^ik_i+m^2}\;\;, \hspace{3ex}
 \underline{k}=(k_0= -\omega_{\vec k},
k_i, K_m) \;\;,
\eeq
and the equal time commutator is 
\bea
[ \underline{a}(k_i, K_m)\, , \,
   \underline{a}^\dagger(k_i^\prime, K_m^\prime) ]
&=&    \frac{1}{ \sqrt{-G}}
\delta^{(p)}(k_i-k_i^\prime)\delta^{(p^\prime-p)}(K_m-K^\prime_m) \;\;\;.
\eea
  A factorized expression is also permitted and we may write
\bea
   \underline{a}(k_i, K_m)     &=&   a(k_i) \alpha(K_m)  \;\;,
   \nonumber \\
\lbrack a(k_i), a^\dagger(k^\prime_i) 
\rbrack &=& \delta^{(p)}(k_i-k_i^\prime)
    \frac{1}{ \sqrt{-G}}  \;\;, \nonumber\\
\lbrack \alpha(K_m), \alpha^\dagger(K^\prime_m) \rbrack &=&
\delta^{(p^\prime-p)}(K_m-K^\prime_m) \;\;.
\eea
Similar expressions hold for $\underline{b}(k_i, K_m)$. We will later use
\beq
\label{eq:exp2}
\phi(x^m) \equiv \frac{1}{(2\pi)^{(p^\prime-p)/2}}\int d^{p^\prime-p}K_m
  \left( \alpha(K_m)e^{i K
  \ppdot x}+\beta^\dagger(K_m)e^{-i K \ppdot x} \right) \;\;.
\eeq
  Reflecting this factorization, we designate Fock space
 assoicated with   $a(k_\mu), b(k_\mu)$ by tach and
 the one with $\alpha(K_m), \beta(K_m)$ by sol.
 Photon Fock space is denoted by vec.

 It is clear that eq. (\ref{eq:fieldeq}) permits an arbitrary field
configuration depending only upon $x^m$ as a solution and this is reflected 
in the expansion  eq. (\ref{eq:exp1}) or eq. (\ref{eq:exp2}). 
To say in a little 
different way, time evolution of the 
operator $\Phi(x^\mu, x^m)$ in the interaction picture is
 independent of how it looks in the $x^m$ direction. 
We will take advantage of this fact to 
accomodate the state representing one noncommutative projector
 soliton with momentum $K_m$ shortly. 

In perturbation theory, the scalar field propagator is 
\bea
\label{eq:propagator}
G(x^M, y^M)&=& \langle 0|T\Phi(x^M)
\Phi^\dagger(y^M)|0\rangle \;\;, \nonumber\\
 &=& \Delta_F(x^\mu-y^\mu; m^2)
\delta^{(p^\prime-p)}(x^m-y^m) \;\;, \nonumber \\
\Delta_F(x^\mu-y^\mu; m^2)&=& \int \frac{d^{(p+1)}k}{(2\pi)^{p+1}}
e^{+ik\pdot x}\otimes e^{-ik\pdot y}   \frac{1}{ \sqrt{-G}} 
\frac{-i}{k \pdot k +m^2-i\epsilon}  \;\;, \nonumber \\
\delta^{(p^\prime-p)}(x^m-y^m)&=& \int 
\frac{d^{(p^\prime-p)} K}{(2\pi)^{p^\prime-p}}
e^{+i K\ppdot x}\otimes e^{-iK\ppdot y}  \;\;.
\eea
In eq. (\ref{eq:propagator}), $x^M$ and $y^M$ are two separate sets
 of integration variables and therefore represent two 
copies of noncommutative algebra. The delta function acts on
 this tensor product space.
Going to the momentum space, one can check 
\begin{equation}
g(x)\ast \int d^{(p^\prime-p)}y\delta^{(p^\prime-p)}(x^m-y^m)\ast 
f(y^m)=g(x^m)\ast f(x^m) ;\;,
\end{equation}
 which means
\begin{equation}
 g(\hat{x}) Tr_{\hat{y}}\hat{\delta}(\hat{x}, 
\hat{y})f(\hat{y})=  g(\hat{x}) f(\hat{x})
 \;\;,
\end{equation}
as it should be.
The Feynmann propagator $\Delta_F(x^\mu-y^\mu)$ in the noncommutative
 space should be understood in the same way. 
  We leave  the discussion on  noncommutative gauge fields to \cite{NCY}.

The interaction Lagrangian ${\cal L}_{\rm int}(\Phi, A_M)$ obtained from
 eq. (\ref{eq:leea}) reads
\bea
{\cal L}_{\rm int}(\Phi, A_M)&=& \frac{1}{2}\Phi^\dagger\ast 
F_{mn}J^{mn}\ast\Phi
-i \Phi^\dagger 
\left( \ast A_\mu\ast  \stackrel{\rightarrow}{\partial}^\mu
  -  \stackrel{\leftarrow}{\partial}^\mu \ast A_\mu\ast 
 \right) \Phi \nonumber\\
& & - \Phi^\dagger\ast A_\mu\ast A^\mu \ast 
\Phi +i[A_M, A_N]_\ast \ast \partial^MA^N \nonumber\\
& &+\frac{1}{4}[A_M, A_N]_\ast \ast [A^M, A^N]_\ast  \;\;.
\eea
 We first determine the  momentum space wave functions  of the initial 
and final states by demanding that the three point amplitude
agree with the string calculation given by eq. (\ref{eq:lima3}). 
Let the initial state associated 
with the scalar field carrying momentum ($k_{1\mu}, K^{(i)}_m=0$) be 
\beq
 |~i~\rangle = |k_{1\mu}, K^{(i)}_{m}=0 \rangle 
             \otimes  \mid k_{3M},  \zeta_{3M}, \cdots, k_{NM},
  \zeta_{NM} \rangle_{\rm  vec} \;\;,
\eeq
and the final state be
\beq
  |~f~\rangle = | -k_{2\mu}, -K^{(f)}_m \rangle 
                \otimes  |\,0 \, \rangle_{\rm vec} \;\;,
\eeq
where
\beq
 | k_\mu,K_m\rangle \equiv  |~k_\mu~\rangle_{\rm tach}
                    \otimes | K_m \rangle \! \rangle_{\rm sol}~,
 \quad |~K_m~\rangle \! \rangle_{\rm   sol}=
u(K_m)\alpha^\dagger(K_m)  |\,0\,\rangle_{\rm  sol} \;\;.
\eeq

The N-point amplitude is
\beq
  {\cal A}_N=\int d^{(p^\prime-p)}K_m \; 
   \langle~f~| \hat{\cal S} |~i~\rangle\;\;,
\eeq
with
\beq
\hat{\cal S}=  T \exp \left[ i\int d^{(p^\prime+1)}x^M 
\sqrt{-G} \, {\cal L}_{\rm int}(\Phi, A_M) \right]\;\;.
\eeq
It is elementary to compute the three point tree amplitude 
 from ${\cal L}_{\rm int}(\Phi, A_M)$ :
\bea
 {\cal A}_3 \!\!\! &=& \!\!\!
  i \int d^{(p^\prime-p)}K_m^{(f)}
    \int d^{(p^\prime+1)}x^M \sqrt{-G} \,
    {}_{\rm  sol}\langle \! \langle - K^{(f)}_m |
      \otimes {_{\rm tach}\langle}-k_{2\mu} |  \nonumber\\
  &&  \left\{ \frac{1}{2}\Phi^\dagger
      \ast {_{\rm vec}\langle}\,0\,|   F_{mn}J^{mn}
                         |k_{3M},  \zeta_{3M} \rangle_{\rm vec}
      \ast \Phi   \right. \nonumber\\
  && \hspace{-3em} \left. -i\Phi^\dagger  
\left( \ast {_{\rm   vec} \langle}\,0\,|
     A_\mu |k_{3M},  \zeta_{3M} \rangle_{\rm vec}\ast 
 \stackrel{\rightarrow}{\partial}^\mu
 -\stackrel{\leftarrow}{\partial}^\mu 
\ast {_{\rm   vec} \langle}\,0\,| A_\mu |k_{3M},  \zeta_{3M}
 \rangle_{\rm vec}\ast  \right)
 \Phi  \right\}  |k_{1\mu}  \rangle_{\rm tach}
\otimes | \, 0 \, \rangle \! \rangle_{\rm  sol} \nonumber\\
 &=& i (2\pi)^{p^\prime+1}\delta^{(p+1)} 
\left(\sum^3_{a=1}k_{a\mu} \right)
  e^{\frac{i}{2}\theta^{\mu\nu}k_{1\mu} k_{2\nu}}
   u^\ast(k_{3m})u(K^{(i)}_m=0)
   \prod_{a=1,2}
    \frac{1}{ \sqrt{(2\pi)^{p^{\prime}}2\omega_{\vec{k}_{a}}} }
\nonumber \\ 
 && \frac{1}{\sqrt{-G}}
     \left( \frac{1}{2}{_{\rm vec}\langle} \,0 \, |
       F_{mn}(0)J^{mn}  |k_{3M}, \zeta_{3M} \rangle_{\rm vec}
 + {_{\rm vec}\langle} \, 0 \, | A(0) \pdot (k_{1}-k_{2})
   | k_{3M},  \zeta_{3M} \rangle_{\rm vec}  \right) \nonumber \\
 &=& -i \left( \frac{1}{\sqrt{-G}} \right)^{2}
     (2\pi)^{p+1}\delta^{(p+1)} \left( \sum^3_{a=1}k_{a\mu} \right) 
     \exp\left(\frac{i}{2}\theta^{\mu\nu}k_{1\mu} k_{2\nu}\right)
       u^\ast(k_{3m})u(0)   \\
  &&  \prod_{a=1,2}\frac{1}{ \sqrt{(2\pi)^{p}2\omega_{{\vec k}_{a}}} }
      \frac{1}{\sqrt{(2\pi)^{p^{\prime}} 2|{\vec k}_{3}|}}
  \left( (k_{2}-k_{1})\prdot \zeta_{3}-ik_{3}\ppdot J\zeta_{3}  \right)
 \nonumber  \;\;.
\label{eq:a3leea}
\eea
Eq.(\ref{eq:lima3}) from string theory and eq.(\ref{eq:a3leea})  computed from 
 eq.(\ref{eq:leea}) agree completely provided 
\beq
\label{eq:id}
u^\ast(k_m)u(0)= D(k_m) \;\;,  \;\;\;{\rm or} \;\;  u(k_m)= D(k_m) \;.
\eeq
  The  momentum space wave function in soliton sector is identified with
  the g.d.f. and hence is equal to Fourier image of the distribution of
  the noncommutative soliton.

   In the alternative point of view discussed in the end of section $2$,
 the intial and final
 states are prepared as
\beq
 \mid u \rangle\rangle_{\rm   sol}  \equiv  \int  d^{(p^\prime-p)}K_m
   \mid K_m\rangle\rangle_{\rm   sol}=   \int  d^{(p^\prime-p)}K_m
u(K_m)\alpha^\dagger(K_m)  \mid 0\rangle_{\rm  sol} \;\;.
\eeq
  The projector soliton configuration is generated by
\beq
   \phi(x^m) \mid u \rangle\rangle_{\rm   sol} \;\;.
\eeq
  This time, our calculation is different from the one at
  eq. (\ref{eq:a3leea}) in that we must integrate over the initial
  soliton momentum $K_m^{(i)}$ as well. We find that the following
  convolution property holds provided  eq. (\ref{eq:id}) is satisfied:
\beqn
\int dK_{m}^{(i)} e^{ +\frac{i}{2} \theta^{mn} K_{m}^{(i)} P_{n} }
u^{\ast} \left( P + K^{(i)} \right) u \left( K^{(i)} \right)
  = u \left( P \right)  \;\;.
\eeqn
The two points of view to the scattering process 
 thus give the same answer.
We have checked that this conclusion also holds for those properties
 other than eq. (\ref{eq:a3leea}) which will be discussed
 in the remainder of this section.

We now proceed to see that the N-point tree amplitude
 obtained from this field
 theory contains the g.d.f. whose argument is the total momentum. 
The amplitude is given by 
\bea
  {\cal A}_N &\sim& \int d^{(p^\prime-p)}K_m^{(f)}
 {_{\rm sol}\langle\langle}-K^{(f)}_m
  \mid \otimes
{_{\rm tach}\langle}-k_{2\mu}|\otimes {_{\rm vec}\langle}0 \mid 
 T \frac{i^{(N-2)} }{(N-2)!}  \\ 
  & &  \hspace{-3em} \left( \int d^{(p^\prime+1)}x^M {\cal L}_{\rm int}
(\Phi, A_M) \right)^{N-2}
 \mid   k_{3M},  \zeta_{3M}, \cdots
k_{NM} ,  \zeta_{NM}   \rangle_{\rm vec} \otimes \mid k_{1\mu} 
\rangle_{\rm tach} \otimes 
\mid  K^{(i)}_m=0 \rangle\rangle_{\rm sol} \;\;\;.   \nonumber 
\eea
 Let us imagine that, in this expression, we first
 carry out the Wick
 contractions and  compute the expectation value for
 the part associated with the vector Fock space. 
To each of the Feynman diagrams generated by this, 
the net effect to the scalar part of the Fock space
 is that there are $L$ vector lines carrying momenta
 $q_{aM},\;$  $a=3 \sim L$
  which are attached to the scalar line. These momenta satisfy
\beq
\sum^L_{a=3}q_{aM}=\sum^N_{a=3}k_{aM} \;\;.
\eeq
The expression for the scalar part of the Fock space coming from this
 Feynman diagram is proportional to
\bea
\label{eq:scalarfock}
  & & \int d^{(p^\prime-p)}K^{(f)}_m {_{\rm sol}\langle\langle}-K^{(f)}_m 
 \mid \otimes
{_{\rm tach}\langle}-k_{2\mu}  \mid   \nonumber \\
  & & T \prod^L_{a=3} \left( \int d^{(p^\prime+1)}x^{M}_a
\Phi^\dagger(x^{M}_a)\ast e^{iq_{a}\prdot x_{a}}\ast \Phi(x^{M}_a) \right)
  \mid k_{1\mu} \rangle_{\rm tach}
\otimes   \mid K^{(i)}_m=0 \rangle \rangle_{\rm sol} \;\;.
\eea
Carrying out the Wick contractions and using the propagator 
(eq. (\ref{eq:propagator})), 
 we find that eq. (\ref{eq:scalarfock}) in turn  contains
the following factor residing in the soliton sector: 
\beqn
   \int d^{(p^\prime-p)}K^{(f)}_m 
\prod^L_{a=3} \left( \int d^{(p^\prime-p)}x^{m}_a \right)
{}_{\rm sol}\langle\langle -K^{(f)}_m  \mid \phi^\dagger(x^{m}_3)\ast 
e^{iq_{3}\ppdot x_{3}}
\ast \delta^{(p^\prime-p)}(x_{3}-x_{4}) \nonumber \\
  \ast  e^{ iq_{4} \ppdot  x_{4}} \ast \cdots 
 \ast \delta^{(p^\prime-p)}(x_{L-1}-x_{L})\ast e^{ iq_{L}\ppdot x_{L}}
 \ast \phi(x^{m}_L)|K^{(i)}_m=0 \rangle \rangle_{\rm sol} \;\;\;. 
\eeqn
  Thanks to the delta function propagator, this equals
\beqn
   &=& \int d^{(p^\prime-p)}K^{(f)}_m \int d^{(p^\prime-p)}x^{m}
   \exp  \left( \frac{i}{2} \displaystyle{ 
\sum_{ {\scriptstyle a,b=3} \atop{\scriptstyle a < b} }^{L} 
\sum_{m, n=  p+1}^{p^{\prime}}
  \theta^{mn} q_{a m} q_{b n} }   \right)   \nonumber\\
   & &{_{\rm sol}\langle\langle}-K^{(f)}_m|\phi^\dagger(x^{m})
  \ast   \exp  \left( {i  \left( {\displaystyle \sum^L_{a=3}}q_{a}
 \right) \ppdot x } 
 \right)  \ast  \phi(x^m)|
  K^{(i)}_m=0\rangle\rangle_{\rm sol} \nonumber \\
 &=& \int d^{(p^\prime-p)}K^{(f)}_m   \delta^{(p^\prime-p)} \left( 
K^{(f)}_m  + (\sum^L_{a=3}q_{a}) \right)
 u^\ast ( - K^{(f)}_m)u(0)
  \exp  \left( \frac{i}{2} \displaystyle{ \sum_{   {\scriptstyle a,b=3} 
 \atop{\scriptstyle a < b} }^{L}
 \sum_{m, n= p+1 }^{p^{\prime}}
  \theta^{mn} q_{am} q_{bn} }     \right)
 \nonumber \\
 &=& D \left(\sum^L_{a=3}k_{am} \right) \exp  \left(  \frac{i}{2}
  \displaystyle{  \sum_{  {\scriptstyle a,b=3}  
 \atop{\scriptstyle a < b}}^{L}
 \sum_{m, n=p+1}^{p^{\prime}}
  \theta^{mn} q_{am} q_{bn} }  \right)    \;\;\;.
\eeqn 
 This completes the demonstration.

Finally, let us check that the tree four point amplitude (the pole part)
 computed from $S$ in 
fact agrees with string answer.
The field theory amplitude is
\begin{eqnarray}
 \lefteqn{
{\cal A}_{4} = \int d^{(p^{\prime}-p)} K_{m}^{f} \,\,\,
 {}_{\rm sol}\langle\!\langle -K_{m}^{(f)} |  \otimes
  {}_{\rm tach} \langle -k_{2\mu} | \otimes {}_{\rm vec} \langle 0| }
    \\  
 && T \left\{ \frac{1}{2!}\,
     i\int d^{p^{\prime}+1}x_{1} \sqrt{-G}{\cal L}_{\rm int}(x_{1})
     i\int d^{p^{\prime}+1}x_{2} \sqrt{-G}{\cal L}_{\rm int}(x_{2})
      \right\}  \nonumber  \\
 &&  |k_{3M}, \zeta_{3M}, k_{4M}, \zeta_{4M} \rangle_{\rm vec} \otimes 
     |k_{1\mu}\rangle \otimes 
     |K_{m}^{(i)}=0\rangle\!\rangle_{\rm sol}~.  \nonumber
\end{eqnarray}
After the Wick contraction and the position space integration,
we find
\begin{eqnarray}
 {\cal A}_{4} \!\!\! &=&\!\!\!
   (2\pi)^{p+1} \delta^{p+1}\left(\sum_{a=1}^{4}k_{a\mu}\right)
     \, \exp\left(\frac{i}{2}\theta^{\mu\nu}k_{1\mu}k_{2\nu}\right)
     u^{\ast}(k_{3M}+k_{4M}) \, u(0)
 \nonumber\\
 && \left(\frac{1}{\sqrt{-G}} \right)^{3} 
   \prod_{a=1,2}
    \frac{1}{\sqrt{ (2\pi)^{p}2\omega_{\vec{k}_{a}} } }
     \prod_{b=3,4}
     \frac{1}{\sqrt{ (2\pi)^{p^{\prime}} |\vec{k}_{b}|} }
       \left( {\bf a}^{(t,u)}_{4} + {\bf a}^{(s)}_{4}   \right)~,
\end{eqnarray}
where
\begin{eqnarray}
{\bf a}_{4}^{(t,u)}\!\!\!&=&\!\!\!
     \frac{-i}{t-m^{2}}
    \left\{ (k_{2}-(k_{1}+k_{4}))\pdot \zeta_{3}-ik_{3}\ppdot J\zeta_{3}
    \right\} \nonumber\\
  && \hspace{4em}
    \left\{(k_{2}+k_{3})-k_{1}) \pdot \zeta_{4}
      {} -ik_{4} \ppdot J\zeta_{4} \right\} \,
      \exp\left(\frac{i}{2}\theta^{MN}k_{3M}k_{4N}\right)
     \nonumber \\
  &&+\left(k_{3}\leftrightarrow k_{4};
           \zeta_{3}\leftrightarrow \zeta_{4} \right)~, \\
{\bf a}_{4}^{(s)} \!\!\! &=& \!\!\!
    \frac{-i}{s} \,\,
    2\left[ \left( (k_{2}-k_{1})\pdot \zeta_{3}
           {}-i(k_{3}+k_{4})\ppdot J\zeta_{3}\right)
                k_{3}\prdot \zeta_{4} \right. \nonumber\\
  && \hspace{3em} - \left( (k_{2}-k_{1}) \pdot \zeta_{4}
           {}-i(k_{3}+k_{4}) \ppdot J \zeta_{4} \right)
                k_{4} \prdot \zeta_{3} \nonumber \\
  && \hspace{3em}\left.
       + \left( \frac{1}{2} (k_{3}-k_{4}) \pdot (k_{1}-k_{2})
           {} -ik_{3} \ppdot J k_{4} \right) 
           \zeta_{3} \prdot \zeta_{4} \right]
           \exp\left( \frac{i}{2}\theta^{MN}k_{3M}k_{4N}\right)
           \nonumber\\
   && + \left( k_{3} \leftrightarrow k_{4};
               \zeta_{3} \leftrightarrow \zeta_{4} \right)~.
\end{eqnarray}
  This expression agrees with   eq.(\ref{eq:st4ampstring}).

\section{Discussion and outlook}
 
 We have seen that perturbation theories of two different kinds in fact
  agree.
 We now make several remarks which are more speculative in nature.
 Our identification at eq. (\ref{eq:id}) tells that the initial and
 the final state wave
 functions in position space obey  equation of motion of
 pure polynomial
 $\phi^{3}$ theory.
  This may indicate that initial/final state interactions are governed
 by $\phi^{3}$ dynamics.
 This point, together with the delta function propagator
 in perturbation theory, is in fact reminiscent 
 of the pregeometric nature of string theory. In perturbation theory,
 we of course  find
 no reason why the wave function of this form has appeared. 
Nonperturbative treatment of the $x_{m}$
  part of the scalar field $\Phi(x_{\mu}, x_{m})$ as
 a soliton operator may
 lead us somewhere beyond what we have accomplished in this paper.
 Finally, tachyon is still present in the spectrum. The system must find
  its ultimate stability. Our preliminary investigation shows
 relevance of a
 noncommutative soliton of a different kind, which is somewhat similar to
  the one of \cite{GN}.

\section{Acknowledgements}

 We would like to thank Kentaro Hori, Tsunehide Kuroki,
 Seiji Terashima and Takeshi Sato
 for useful discussion on this subject.
 H.I. and K.M. are also grateful to the hospitality of
 the workshop ``Summer Institute 2000" at Yamanashi, Japan,
 where a part of this work was carried out.


\end{document}